\documentclass[11pt,a4paper]{article}

\usepackage{amssymb}
\usepackage{epsfig}

\begin{document}

\title{Reissner-Nordstr\"{o}m black hole lensing}
\author{Ernesto F. Eiroa$^{1,}$\thanks{
e-mail: eiroa@iafe.uba.ar}, Gustavo E. Romero$^{2,}$\thanks{
e-mail: romero@irma.iar.unlp.edu.ar. Member of CONICET}, and Diego F. Torres$
^{3,}$\thanks{
e-mail: dtorres@princeton.edu} \\
{\small $^1$ Instituto de Astronom\'{\i}a y F\'{\i}sica del Espacio, C.C.
67, Suc. 28, 1428, Buenos Aires, Argentina}\\
{\small $^2$Instituto Argentino de Radioastronom\'{\i}a, C.C.5, 1894 Villa
Elisa, Buenos Aires, Argentina}\\
{\small $^3$ Physics Department, Princeton University, NJ 08544, USA}}
\maketitle
\date{}

\begin{abstract}
In this paper we study the strong gravitational lensing scenario
where the lens is a Reissner-Nordstr\"{o}m black hole. We obtain
the basic equations and show that, as in the case of Schwarzschild
black hole, besides the primary and secondary images, two infinite
sets of relativistic images are formed. We find analytical
expressions for the positions and amplifications of the
relativistic images. The formalism is applied to the case of a
low-mass black hole placed at the galactic halo.
\end{abstract}


\section{Introduction}

The theory of General Relativity predicts the deflection of light
in presence of a mass distribution. A. Einstein in 1936
\cite{einstein} noted that an image due to the deflection of the
light of a background star by another star can have a great
magnification if the observer, the lens, and the source are highly
aligned. He also pointed out that the angular separation of the
images was too small to be resolved by the optical
telescopes available at that time.\\

It was the discovering of quasars in 1963 which opened the possibility of
really observing gravitational lensing effects. Quasars are very bright
objects, located at cosmological distances, and have a central compact optical
emitting region. When a galaxy is interposed in the line of sight to them, 
the resulting gravitational magnification can be large and the images are well
separated in some particular cases. In 1979 the first example of
gravitational lensing was discovered (the quasar QSO 0957+561 A,B). \\

The weak field theory of gravitational lensing, developed, among others, by
Y. G. Klimov, S. Liebes, S. Refsdal, R. R. Bourassa, and R. Kantowski, has
been successful in explaining the astronomical observations up to now. This
theory is based on a first order expansion of the small deflection angle
(for a detailed treatment see \cite{schneider}, and references therein). \\

When the lens is a very compact object (e.g. a black hole) the
weak field approximation is no longer valid. Virbhadra and Ellis
recently studied the strong field situation \cite{virbha}. They
obtained the lens equation using an asymptotically flat background
metric and analyzed the lensing by a Schwarzchild black hole in
the center of the Galaxy using numerical methods. Besides the
primary and secondary images, they found that there exist two
infinite sets of faint relativistic images. Fritelli et al. 
\cite{fritelli} found an exact lens
equation without any reference to a background metric and compared
their results with those of Virbhadra and Ellis for the
Schwarzchild black hole case. Bozza et al. \cite{bozza} obtained
analytical expressions for the positions and magnification of the
relativistic images using the strong
field limit approximation.\\

In this paper we study the lensing situation when the lens is a
slowly rotating Kerr-Newman black hole. The introduction of
charged bodies in the strong-field lensing theory is justified
since charged black holes are thought to be final result of the
catastrophic collapse of very massive ($ M>35$ $M_{\odot}$)
magnetized stars. Although selective accretion from the
surroundings would neutralize the charged black hole if it is
located in a high-density medium, there remains the possibility
that if the Kerr-Newman hole is surrounded by a co-rotating,
opposite charged magnetosphere, it might not discharge so quickly.
Such a configuration would present zero net charge from infinity
and consequently it could survive for a significant time span
($10^3-10^5$ yr) if located in a low density environment \cite
{punsly98}. Moreover, the magnetosphere would have observational
effects due to particle acceleration in electrostatic polar gaps,
similar to those presented by pulsars \cite{punsly98}. The
relativistic wind created by the Kerr-Newman black hole could be
responsible for detectable gamma-ray emission \cite{Punsly:2000xb}.
Hence, the study of other possible observational signatures
from charged black holes presents particular interest.\\

In Sec. 2 of this paper we introduce the basic equations using the
Reissner-Nordstr\"om's metric for the black hole and a flat background
metric. The Reissner-Nordstr\"om case can be used also to study slowly
rotating Kerr-Newman black holes. Highly rotating objects break the
spherical symmetry introducing unnecessary complications in the lensing
calculations, which do not lead to qualitatively different results. In Sec.
3 we use the strong field limit approximation to obtain analytical
expressions for the positions and amplifications of the relativistic images.
In Sec. 4 we calculate the positions and magnifications of the primary and
secondary images using the weak field approximation. In Sec. 5 we apply the
formalism to a small black hole (7 solar masses) in the galactic halo. We
close with Sec. 6, where some conclusions are drawn.

\section{Basic equations}

In this paper we use geometrized units (speed of light in vacuum $c=1$ and
gravitational constant $G=1$).\\

Black holes are characterized uniquely by $M$ (mass), $Q$ (charge) and $S$
(intrinsic angular momentum) \cite{misner}. Written in the $t$, $r$, $\theta
$,$\varphi $ coordinates of Boyer and Lindquist, the Kerr-Newman geometry
has the form:
\begin{equation}
ds^{2}=\frac{-\Delta }{\rho ^{2}}(dt-a\sin ^{2}\theta d\varphi )^{2}+\frac{
\sin ^{2}\theta }{\rho ^{2}}[(r^{2}+a^{2})d\varphi -adt]^{2}+\frac{\rho ^{2}
}{\Delta }dr^{2}+\rho ^{2}d\theta ^{2},  \label{1}
\end{equation}
where
\begin{equation}
\Delta =r^{2}-2Mr+a^{2}+Q^{2},  \label{1a}
\end{equation}
\begin{equation}
\rho ^{2}=r^{2}+a^{2}\cos ^{2}\theta ,  \label{1b}
\end{equation}
\begin{equation}
a=\frac{S}{M}.  \label{1c}
\end{equation}

The Kerr-Newman geometry is axially symmetric around z axis. The
horizon of events is placed at
\begin{equation}
r_{\mathrm{H}}=r_{+}=M+\sqrt{M^{2}-Q^{2}-a^{2}}.  \label{2}
\end{equation}

A non-rotating black hole corresponds to an isotropic black hole
with charge Q. The metric is the Reissner-Nordstr\"{o}m's one:
\begin{equation}
ds^{2}=-\left( 1-\frac{2M}{r}+\frac{Q^{2}}{r^{2}}\right) dt^{2}+r^{2}\left(
\sin ^{2}\theta d\varphi ^{2}+d\theta ^{2}\right) +\left( 1-\frac{2M}{r}+
\frac{Q^{2}}{r^{2}}\right) ^{-1}dr^{2}.  \label{3}
\end{equation}

The horizon is located at
\begin{equation}
r_{\mathrm{H}}=M+\sqrt{M^{2}-Q^{2}},  \label{4}
\end{equation}
and the photon sphere radius is at
\begin{equation}
r_{\mathrm{ps}}=\frac{3}{2}M\left( 1+\sqrt{1-\frac{8}{9}\frac{Q^{2}}{M^{2}}}
\right) .  \label{5}
\end{equation}

\begin{figure}[t]
\vspace{-3cm}
\par
\begin{center}
\includegraphics[width=10cm,height=14cm]{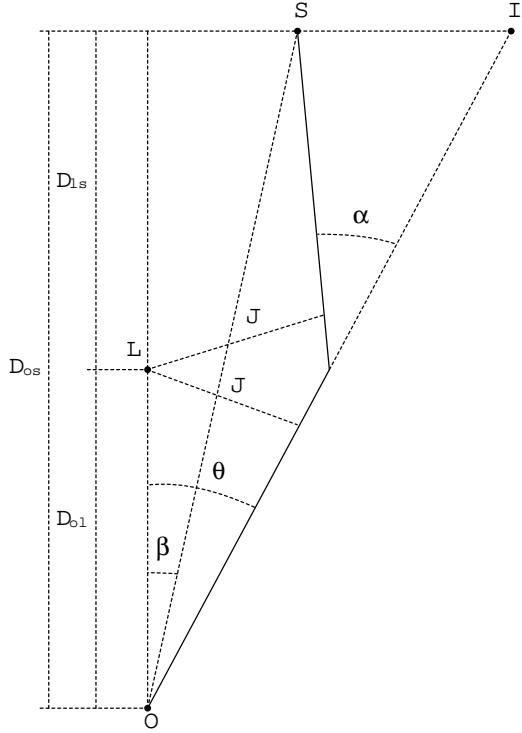} \vspace{-1.5cm}
\end{center}
\caption{ Lens diagram. The observer (O), the lens (L), the source (S) and
the image (I) positions are shown. $D_{\mathrm{ol}}$, $D_{\mathrm{os}}$, $D_{
\mathrm{ls}}$ are, respectively, the observer-lens, the observer-source and
the lens-source distances. $\alpha$ is the deflection angle and $J$ is the
impact parameter.}
\label{rnf1}
\end{figure}

We study the lens situation shown in Fig. \ref{rnf1}. The
background space-time is considered asymptotically flat, with the
observer and the source immersed in the flat space-time region,
which can be embedded, if necessary, in a Robertson-Walker
expanding Universe. From Fig. 1 we see that the lens equation can
be expressed \cite{virbha}
\begin{equation}
\tan \beta =\tan \theta -\frac{D_{\mathrm{ls}}}{D_{\mathrm{os}}}\left( \tan
\theta + \tan(\alpha -\theta )\right) ,  \label{6}
\end{equation}
where $\beta $ and $\theta $ are respectively the angular source and image
positions and $\alpha $ is the deflection angle due to the black hole.
\\

Following Sec. 8.5 of \cite{weinberg}, we have that the deflection angle for
a light ray is
\begin{equation}
\alpha (r_{0})=2\int_{r_{0}}^{\infty }\frac{dr}{r\sqrt{\left( \frac{r}{r_{0}}
\right) ^{2}\left( 1-\frac{2M}{r_{0}}+\frac{Q^{2}}{r_{0}^{2}}\right) -\left(
1-\frac{2M}{r}+\frac{Q^{2}}{r^{2}}\right) }}-\pi ,  \label{7}
\end{equation}
where $r_{0}$ is the closest distance of approach. The impact parameter is
\begin{equation}
J(r_{0})=r_{0}\left( 1-\frac{2M}{r_{0}}+\frac{Q^{2}}{r_{0}^{2}}\right) ^{-
\frac{1}{2}}.  \label{8a}
\end{equation}

From the lens diagram (Fig. \ref{rnf1}) we see that
\begin{equation}
J(r_{0})=D_{\mathrm{ol}}\sin \theta .  \label{8b}
\end{equation}

Defining the distances and the charge in terms of the Schwarzschild radius $
(2M)$:

\begin{eqnarray}
x=\frac{r}{2M},\hspace{1cm} x_{0}=\frac{r_{0}}{2M}, \hspace{1cm} b=\frac{ J
}{2M},  \nonumber
\end{eqnarray}
\begin{eqnarray}
d_{\mathrm{ol}}=\frac{D_{\mathrm{ol}}}{2M}, \hspace{1cm} d_{\mathrm{os}}=
\frac{D_{\mathrm{os}}}{2M}, \hspace{1cm} d_{\mathrm{ls}}=\frac{D_{\mathrm{ls
}}}{2M},  \nonumber
\end{eqnarray}
\begin{eqnarray}
q=\frac{Q}{2M},  \nonumber
\end{eqnarray}

we have that
\begin{equation}
\alpha (x_{0})=2\int_{x_{0}}^{\infty }\frac{dx}{x\sqrt{\left( \frac{x}{x_{0}}
\right) ^{2}\left( 1-\frac{1}{x_{0}}+\frac{q^{2}}{x_{0}^{2}}\right) -\left(
1-\frac{1}{x}+\frac{q^{2}}{x^{2}}\right) }}-\pi ,  \label{10}
\end{equation}
\begin{equation}
b(x_{0})=x_{0}\left( 1-\frac{1}{x_{0}}+\frac{q^{2}}{x_{0}^{2}}\right) ^{-
\frac{1}{2}}=d_{\mathrm{ol}}\sin \theta ,  \label{11}
\end{equation}
\begin{equation}
x_{\mathrm{H}}=\frac{1}{2}+\sqrt{\frac{1}{4}-q^{2}},  \label{12}
\end{equation}
\begin{equation}
x_{\mathrm{ps}}=\frac{3}{4}\left( 1+\sqrt{1-\frac{32}{9}q^{2}}\right) .
\label{13}
\end{equation}

\begin{figure}[t]
\vspace{-3cm}
\par
\begin{center}
\includegraphics[width=10cm,height=14cm]{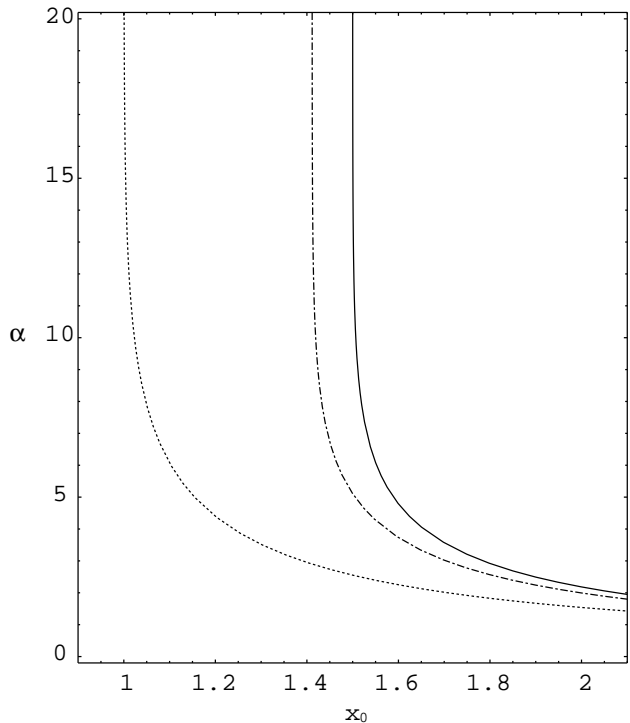} \vspace{-1.5cm}
\end{center}
\caption{ Deflection angle $\alpha $ (in radians) plotted as a
function of the closest approach distance $x_{0}$. The solid line
corresponds to $Q=0$, the dash-dot line to $\left| Q\right| =0.5M$
and the dot line to $\left| Q\right| =M$ . In each curve the
vertical asymptote is placed at $x_{0}=x_{ \mathrm{ps}}$.}
\label{rnf2}
\end{figure}

\begin{figure}[t]
\vspace{-1cm}
\par
\begin{center}
\includegraphics[width=10cm,height=14cm]{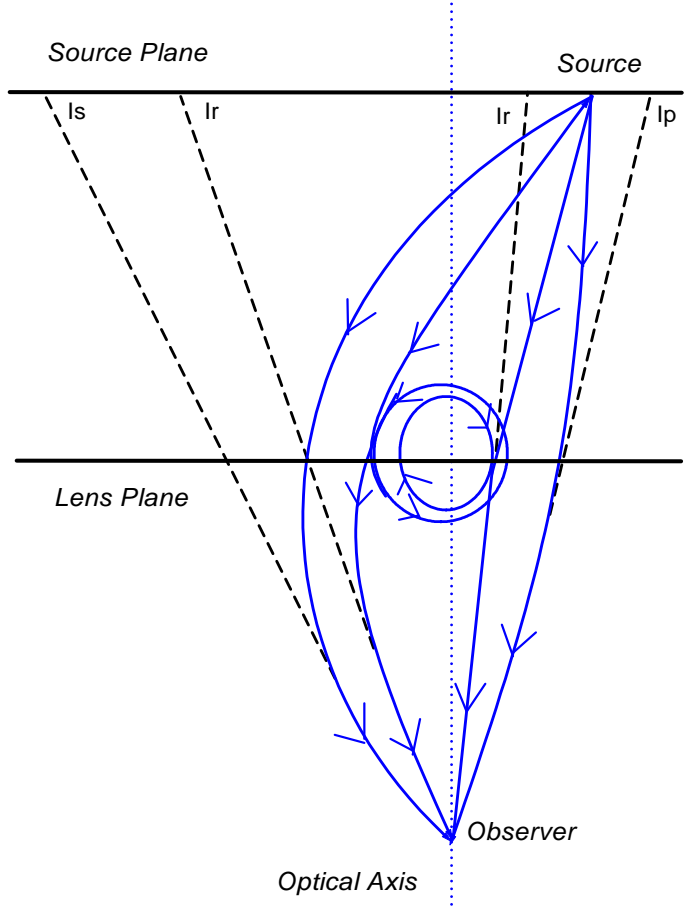} \vspace{0cm}
\end{center}
\caption{The primary($I_{\mathrm{p}}$) and secondary ($I_{\mathrm{s}}$)
images are formed when the deflection angle is (in modulus) smaller than $
2\pi $. The relativistic images ($I_{\mathrm{R}}$) are obtained when the
deflection angle is (in modulus) greater than $2\pi $ (the first with
clockwise winding and the first with counter-clockwise winding are shown). }
\label{rnf3}
\end{figure}

The deflection angle $\alpha $ for a light ray passing at the
right of the black hole is plotted as a function of the closest
approach distance $x_{0}$ in Fig. \ref{rnf2}. We see that when
$x_{0}$ takes values near $x_{\mathrm{ ps }}$ the angle of
deflection $\alpha $ is greater than $2\pi $ , so the light ray
can take several turns around the black hole before reaching the
observer. In this way, besides the primary and secondary images
(with $ \left| \alpha \right| <2\pi $), we have two infinite sets
of relativistic images, one produced by clockwise winding around
the black hole ($\alpha >0$ ) and the other by counter-clockwise
winding ($\alpha <0$). These images are located, respectively, at
the same side and at the opposite side of the
source. This is qualitatively shown in Fig. \ref{rnf3}. \\

For a given source position $\beta $ , we must solve Eq. (\ref{6}) with Eqs.
(\ref{10}) and (\ref{11}) to obtain the positions of the images. The
trascendental equation (\ref{6}) is hard to solve even numerically. \\

The magnification for a circular symmetric lens is given by
\begin{equation}
\mu =\left| \frac{\sin \beta }{\sin \theta }\frac{d\beta }{d\theta }\right|
^{-1}.  \label{21}
\end{equation}

Differentiating both sides of Eq. (\ref{6}) and with some algebra, we have
\begin{equation}
\frac{d\beta }{d\theta }=\left( \frac{\cos \beta }{\cos \theta }\right)
^{2}\left\{ 1-\frac{d_{\mathrm{ls}}}{d_{\mathrm{os}}}\left[ 1+\left( \frac{
\cos \theta }{\cos (\alpha -\theta )}\right) ^{2}\left( \frac{d\alpha }{d\theta }-1 \right) \right] \right\} ,  \label{22}
\end{equation}
where
\begin{equation}
\frac{d\alpha }{d\theta }=\frac{d\alpha }{dx_{0}}\frac{dx_{0}}{d\theta },
\label{23}
\end{equation}
with
\begin{equation}
\frac{d\alpha }{dx_{0}}=\int_{x_{0}}^{\infty }\frac{-x^{4}\left(
x_{0}-2q^{2}\right) +x_{0}^{4}\left( x-2q^{2}\right) }{x_{0}^{5}x^{3}\left[
\left( \frac{x}{x_{0}}\right) ^{2}\left( 1-\frac{1}{x_{0}}+\frac{q^{2}}{
x_{0}^{2}}\right) -\left( 1-\frac{1}{x}+\frac{q^{2}}{x^{2}}\right) \right] ^{
\frac{3}{2}}}dx,  \label{24}
\end{equation}
obtained by differentiating Eq. (\ref{10}) and
\begin{equation}
\frac{dx_{0}}{d\theta }=\frac{x_{0}^{2}\left( 1-\frac{1}{x_{0}}+\frac{q^{2}}{
x_{0}^{2}}\right) ^{\frac{3}{2}}\sqrt{1-\left( \frac{x_{0}}{d_{\mathrm{ol}}}
\right) ^{2}\left( 1-\frac{1}{x_{0}}+\frac{q^{2}}{x_{0}^{2}}\right) ^{-1}}}{
\frac{1}{ 2d_{\mathrm{ol}}}\left( 2x_{0}^{2}-3x_{0}+2q^{2}\right) },
\label{25}
\end{equation}
obtained by differentiating both sides of Eq. (\ref{11}) and doing
some extra algebraic tricks.

\section{Strong field limit}

In this section we shall do some approximations. The first one is
that when the source and the lens are almost aligned we can
replace $\tan \theta $ by $ \theta $ and $\tan \beta $ by $\beta
$. For the relativistic images with clockwise winding of the rays
around the black hole, we can write $\alpha =2 \mathrm{{n}\pi
+\Delta \alpha _{{n}}}$ with n integer and $0<\Delta \alpha _{
\mathrm{n}}<<1,$ so that $\tan (\alpha -\theta )$ can be
approximated by $\Delta \alpha _{\mathrm{n}}-\theta $. As in the
case of Schwarzschild black hole lensing, if a ray of light
emitted by the source $S$ is going to reach the observer after
turning around the black hole, $\alpha$ must be very close to a
multiple of $2 \pi$.

 Then the lens equation takes the form \cite {bozza}:
\begin{equation}
\beta =\theta -\frac{D_{\mathrm{ls}}}{D_{\mathrm{os}}}\Delta \alpha _{
\mathrm{n}}= \theta -\frac{d_{\mathrm{ls}}}{d_{\mathrm{os}}}\Delta \alpha _{
\mathrm{n}},  \label{26}
\end{equation}
and the impact parameter is
\begin{equation}
b\approx \frac{D_{\mathrm{ol}}}{2M}\theta =d_{\mathrm{ol}}\theta .
\label{36}
\end{equation}

The relativistic images are formed when the light rays pass very close to
the photon sphere. So it is convenient to write the closest approach
distance $x_{0}$ as
\begin{equation}
x_{0}=x_{\mathrm{ps}}+\varepsilon ,  \label{27}
\end{equation}
where $0\leq \varepsilon <<1$. Bozza et al \cite{bozza} have shown for the
Schwarzschild black hole that the deflection angle can be approximated by:
\begin{equation}
\alpha =-2\ln \left( \frac{2+\sqrt{3}}{18}\varepsilon \right) -\pi .
\label{28}
\end{equation}

From Fig. \ref{rnf2} we see that the curves representing the
deflection angle have similar form for the different values of
$Q$. Therefore, we shall also look for a similar approximation
\begin{equation}
\alpha =-A\ln (B\varepsilon )-\pi ,  \label{29}
\end{equation}
where $A$ and $B$ are to-be-defined positive numbers which will depend only on $q=Q/2M$.\\

$A$ and $B$ are chosen to satisfy
\begin{equation}
\lim_{x_{0}\rightarrow x_{\mathrm{ps}}}\left( \alpha _{\mathrm{exact}}-
\alpha _{\mathrm{approx.}}\right) =0,  \label{30a}
\end{equation}
with $\alpha _{\mathrm{exact}}$ given by Eq.
(\ref{10})\footnote{The integral of Eq. (\ref{10}) can be
calculated in terms of elliptic integrals and it is done in the
Appendix, Eq. (\ref{a3}).} and $\alpha _{\mathrm{approx.}}$ is
given by Eq. (\ref{29}). Eq. (\ref{30a}) can be written in the
form
\begin{equation}
\lim_{x_{0}\rightarrow x_{\mathrm{ps}}}\left[ \alpha _{\mathrm{approx.}}\left( \frac{\alpha _{\mathrm{exact}}}{\alpha _{\mathrm{approx.}}}-1\right) \right] =0.  \label{30b}
\end{equation}
A necessary (but not sufficient) condition for this limit is:
\begin{equation}
\lim_{x_{0}\rightarrow x_{\mathrm{ps}}}\frac{\alpha _{\mathrm{exact}}}{
\alpha _{\mathrm{approx.}}}=1.  \label{30c}
\end{equation}
As $\alpha \rightarrow \infty$ when $x_{0}\rightarrow x_{\mathrm{ps}}$, using the L'Hospital's rule in the limit of Eq. (\ref{30c}), it is easy to find $A$:
\begin{equation}
A=\lim_{x_{0}\rightarrow x_{\mathrm{ps}}}\left[ -(x_{0}-x_{\mathrm{ps}})
\frac{d\alpha _{\mathrm{exact}}}{dx_{0}}\right] ,  \label{32}
\end{equation}
with $\frac{d\alpha _{\mathrm{exact}}}{dx_{0}}$ given by Eq. (\ref{24})\footnote{ Or, in terms of elliptic integrals, by Eq. (\ref{a11}) of the Appendix.}.

To obtain $B$, we replace Eq.(\ref{29}) in Eq. (\ref{30a}):
\begin{equation}
\lim_{x_{0}\rightarrow x_{\mathrm{ps}}}\left[ \alpha _{\mathrm{exact}}+A\ln B(x_{0}-x_{\rm{ps}})+\pi \right]=0,  \label{33a}
\end{equation}
which, using properties of logarithms can be expressed as
\begin{equation}
\lim_{x_{0}\rightarrow x_{\mathrm{ps}}}A\ln \left[ B(x_{0}-x_{\rm{ps}})\exp \left( \frac{\alpha _{\mathrm{exact}}+\pi }{A}\right) \right] =0,  \label{33b}
\end{equation}
so
\begin{equation}
\lim_{x_{0}\rightarrow x_{\mathrm{ps}}}B(x_{0}-x_{\rm{ps}})\exp \left( \frac{\alpha _{\mathrm{exact}}+\pi }{A}\right) =1,  \label{33c}
\end{equation}
then
\begin{equation}
B=\lim_{x_{0}\rightarrow x_{\mathrm{ps}}}\frac{\exp \left[ \frac{-(\alpha
_{\mathrm{exact}}+\pi )}{A}\right] }{(x_{0}-x_{\mathrm{ps}})}.  \label{33}
\end{equation}
$A$ and $B$ are given in Table 1 for some values of $Q$.

\begin{table}[h]
\caption{Numerical values for the coefficients $A$ and $B$. See
text for explanation. }
\centering
\begin{tabular}{ccccccc}
\hline
$\left| Q\right| $ & $0$ & $0.1M$ & $0.25M$ & $0.5M$ & $0.75M$ & $1M$ \\
\hline
$A$ & $2.00000$ & $2.00224$ & $2.01444$ & $2.06586$ & $2.19737$ & $2.82843$
\\
$B$ & $0.207338$ & $0.207979$ & $0.21147$ & $0.225997$ &
$0.262085$ & $ 0.426782$ \\ \hline
\end{tabular}
\label{rnt1}
\end{table}

\begin{figure}[t]
\vspace{-3cm}
\par
\begin{center}
\includegraphics[width=10cm,height=14cm]{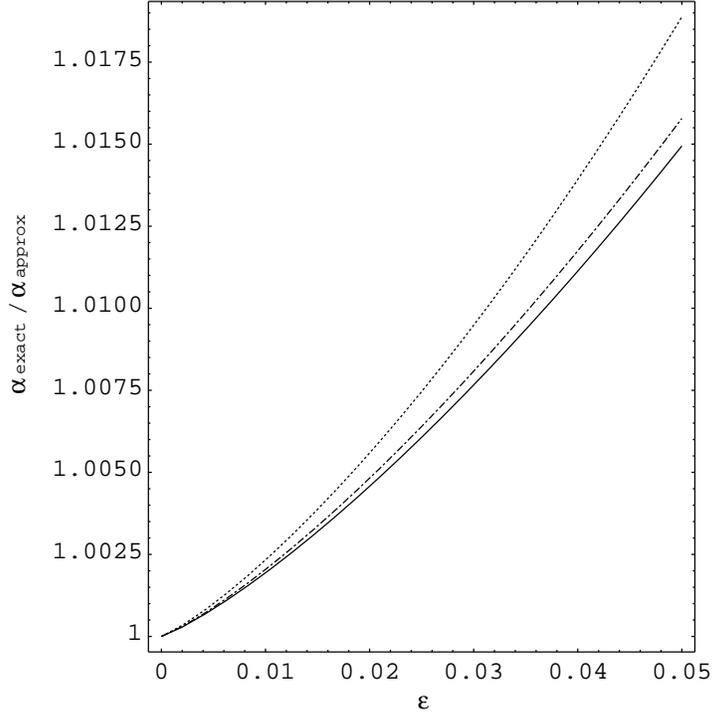} \vspace{-1.5cm}
\end{center}
\caption{The ratio $\alpha _{\mathrm{exact}}/\alpha _{\mathrm{approx.}}$
plotted as a function of $\varepsilon =x_{0}-x_{\mathrm{ps}}$. The solid
line corresponds to $Q=0$, the dash-dot line to $\left| Q\right| =0.5M$ and
the dot line to $\left|Q\right| =M$. }
\label{rnf4}
\end{figure}

The ratio $\alpha _{\mathrm{exact}}/\alpha _{\mathrm{approx.}}$ is plotted
in Fig. \ref{rnf4} for $0\leq \varepsilon \leq 0.05$. The error using $
\alpha _{\mathrm{approx.}}$ is very small, less than 1.5 percent for $\left|
Q\right| \leq 0.5M$ and less than 2 percent for $0.5M\leq \left| Q\right|
\leq M$.\\

The impact parameter given by Eq. (\ref{11}) can be approximated, using a
second order Taylor expansion in $\varepsilon $, by
\begin{equation}
b=C+D\varepsilon ^{2},  \label{34a}
\end{equation}
with
\begin{equation}
C=\frac{\left( 3+\sqrt{9-32q^{2}}\right) ^{2}}{4\sqrt{6-16q^{2}+2\sqrt{
9-32q^{2}}}},  \label{34b}
\end{equation}
and
\begin{equation}
D=\frac{72-256q^{2}}{\left( 9-32q^{2}+\sqrt{9-32q^{2}}\right) \sqrt{
6-16q^{2}+2\sqrt{9-32q^{2}}}},  \label{34c}
\end{equation}
where we have made use of Eqs. (\ref{13}, \ref{27}).
Inverting the Eq. (\ref{34a}) to obtain $\varepsilon $
\begin{equation}
\varepsilon =\sqrt{\frac{b-C}{D}},  \label{35}
\end{equation}
replacing Eqs. (\ref{36}) and (\ref{35}) in Eq. (\ref{29}) we have
\begin{equation}
\alpha \approx -A\ln \left( B\sqrt{\frac{d_{\mathrm{ol}}\theta -C}{D}}
\right) -\pi .  \label{37}
\end{equation}

The position of the n-th relativistic image can be approximated by a first order Taylor expansion around $\alpha =2\rm{n}\pi $
\begin{equation}
\theta _{\mathrm{n}}\approx \theta _{\mathrm{n}}^{0}-\rho _{\mathrm{n}
}\Delta \alpha _{\mathrm{n}},  \label{38}
\end{equation}
with
\begin{equation}
\theta _{\mathrm{n}}^{0}=\theta (\alpha =2n\pi ),  \label{39}
\end{equation}
and
\begin{equation}
\rho _{\mathrm{n}}=\left. -\frac{d\theta }{d\alpha } \right| _{\alpha =2n\pi
}.  \label{40}
\end{equation}
Inverting Eq. (\ref{37}) we have that
\begin{equation}
\theta =\frac{1}{d_{\mathrm{ol}}} \left\{ \frac{D}{B^{2}}
\exp \left[ \frac{-2}{A}(\alpha +\pi) \right] +C \right\} ,  \label{37b}
\end{equation}
then
\begin{equation}
\theta _{\mathrm{n}}^{0}=\frac{1}{d_{\mathrm{ol}}} \left\{ \frac{D}{B^{2}}
\exp \left[ \frac{-2}{A}(2n+1)\pi \right] +C \right\} ,  \label{41}
\end{equation}
and
\begin{equation}
\rho _{\mathrm{n}}=\frac{2D}{d_{\mathrm{ol}}AB^{2}}\exp \left[ \frac{-2}{A}
(2n+1)\pi \right] .  \label{42}
\end{equation}

From Eq. (\ref{38})
\begin{equation}
\Delta \alpha _{\mathrm{n}}\approx \frac{\theta _{\mathrm{n}}-\theta _{
\mathrm{n}}^{0}}{-\rho _{\mathrm{n}}},  \label{43}
\end{equation}
replacing it in the lens equation, Eq. (\ref{26}):
\begin{eqnarray}
\beta & = & \theta _{\rm{n}}+\frac{d_{\rm{ls}}}{d_{\rm{os}}} \frac{\theta _{\rm{n}}-\theta _{\rm{n}}^{0}}{\rho _{\rm{n}}} \nonumber \\
& =  & \left( 1+\frac{d_{\rm{ls}}}{d_{\rm{os}}\rho _{\rm{n}}}\right) \theta _{\rm{n}} +\frac{d_{\rm{ls}}}{d_{\rm{os}}\rho _{\rm{n}}}\theta _{\rm{n}}^{0}.\label{43b}
\end{eqnarray}
Using that $\rho _{\mathrm{n}}\varpropto \frac{1}{d_{\mathrm{ol}}}$, then
$\frac{d_{\mathrm{ls}}}{d_{\mathrm{os}}\rho _{\mathrm{n}}}\varpropto
\frac{d_{\mathrm{ls}}d_{\mathrm{ol}}}{d_{\mathrm{os}}}>>1$, we can neglect
the $1$ inside the parentheses to obtain  the approximate positions of the
relativistic images:
\begin{equation}
\theta _{\mathrm{n}}=\theta _{\mathrm{n}}^{0}+\frac{d_{\mathrm{os}}\rho _{
\mathrm{n}}}{d_{\mathrm{ls}}}\beta .  \label{44a}
\end{equation}

Note that the second term in the last equation is a small correction on $
\theta _{\mathrm{n}}^{0}$. When the source and the lens are aligned, $\beta
=0$ and we obtain the relativistic Einstein rings with angular radius $
\theta _{\mathrm{n}}^{\mathrm{E}}=\theta _{\mathrm{n}}^{0}$.\\

The amplification of the n-th image is given by
\begin{equation}
\mu _{\mathrm{n}}\approx \left| \frac{\beta }{\theta _{\mathrm{n}}}\frac{
d\beta }{d\theta _{\mathrm{n}}}\right| ^{-1},  \label{45}
\end{equation}
which, using Eq. (\ref{44a}), gives
\begin{equation}
\mu _{\mathrm{n}}=\left| \left( \theta _{\mathrm{n}}^{0}+\frac{d_{\mathrm{os}}\rho _{\mathrm{n}}}{d_{\mathrm{ls}}}\beta \right) \frac{d_{\mathrm{os}}\rho _{\mathrm{n}}}{\beta d_{\mathrm{ls}}} \right|,
 \label{45b}
\end{equation}
neglecting the term with $\left( \frac{d_{\mathrm{os}}\rho _{\mathrm{n}}}{d_{
\mathrm{ls}}}\right) ^{2}$ we have
\begin{equation}
\mu _{\mathrm{n}}\approx \frac{1}{\left| \beta \right| }\frac{d_{\mathrm{os}
} }{d_{\mathrm{ls}}} \theta _{\mathrm{n}}^{0}\rho _{\mathrm{n}}. \label{46}
\end{equation}
We see that $\mu _{\mathrm{n}}\propto \frac{d_{0S}}{\beta d_{\mathrm{ls}}d_{
\mathrm{ol}}^{2}}$, so the amplification of the relativistic images is very
small unless the observer, the lens and the source are highly aligned.
\\

With a similar treatment for the images formed with counter-clockwise
winding of light rays around the black hole, the positions of the
relativistic images are
\begin{equation}
\theta _{\mathrm{n}}=-\theta _{\mathrm{n}}^{0}+\frac{d_{\mathrm{os}}\rho _{
\mathrm{n}}}{d_{\mathrm{ls}}}\beta ,  \label{44b}
\end{equation}
and the amplifications are given again by Eq. (\ref{46}), but with opposite
parity.\\

To obtain the total magnification of the relativistic images, we must take
into account both sets of relativistic images, and sum
\begin{equation}
\mu _{\mathrm{R}}=2\sum\limits_{n=1}^{\infty }\mu _{\mathrm{n}}=\frac{2}{
\left| \beta \right| }\frac{d_{\mathrm{os}}}{d_{\mathrm{ls}}}
\sum\limits_{n=1}^{\infty } \theta _{\mathrm{n}}^{0}\rho _{\mathrm{n}
} ,  \label{47a}
\end{equation}
which, using that the sum of a geometrical series is $\sum\limits_{n=1}^{
\infty }a^{\mathrm{n}}=\frac{a}{1-a}$ for $\left| a\right| <1$, gives
\begin{equation}
\mu _{\mathrm{R}}\approx \frac{1}{\left| \beta \right| }\frac{d_{\mathrm{os}
} }{d_{\mathrm{ls}}d_{\mathrm{ol}}^{2}}\frac{4D}{AB^{2}}\left( \frac{D}{B^{2}
} \frac{e^{-12\pi /A}}{1-e^{-8\pi /A}}+C\frac{e^{-6\pi /A}}{1-e^{-4\pi /A}}
\right).  \label{47b}
\end{equation}

\section{Primary and secondary images}

We saw in the last section that relativistic images have a very small
amplification unless the aligment is high. Now we shall obtain the positions
and amplifications for the primary and secondary images in the aproximation
of very small source angle ($\beta $).\\

Very small $\beta $ implies, from the lens equation, that $\alpha $ is very
small too for the primary ($P$) and secondary images ($S$), so we can use
the weak field aproximation to obtain the positions and amplifications. The
first step is to calculate the deflection angle $\alpha $. Expanding the
integrand of Eq. (\ref{10}) and doing the integration we have that
\begin{equation}
\alpha =\frac{2}{x_{0}}+\left[ \left( \frac{15}{16} \pi -1 \right) -
\frac{3}{4}\pi q^{2}\right] \frac{1}{x_{0}^{2}} + \mathcal{O}
\left( \frac{1}{x_{0}^{3}} \right).  \label{48}
\end{equation}
The charge only introduces a small correction in the second order term of
the deflection angle. As usual in the weak field aproximation, we shall use
a first order expansion, taking $\alpha \approx 2/x_{0}$ and $b \approx
x_{0} \approx d_{\mathrm{ol}} \theta $. Introducing $\alpha$ in the lens
equation and inverting it to obtain the positions of the images, we have:
\begin{equation}
\theta _{\mathrm{p}}=\frac{1}{2}\left( \beta +\sqrt{\beta ^{2}+4\theta _{
\mathrm{E}}^{2}}\right) \approx \theta _{\mathrm{E}}+\frac{1}{2}\beta ,
\label{49a}
\end{equation}
\begin{equation}
\theta _{\mathrm{s}}=\frac{1}{2}\left( \beta -\sqrt{\beta ^{2}+4\theta _{
\mathrm{E}}^{2}}\right) \approx -\theta _{\mathrm{E}}+\frac{1}{2}\beta ,
\label{49b}
\end{equation}
where $\theta _{\mathrm{E}} =\sqrt{\frac{2d_{\mathrm{ls}}}{d_{\mathrm{os}}d_{
\mathrm{ol}}}}$ is the Einstein ring angular radius. Using Eq. (\ref{21}),
the amplifications are given by:
\begin{equation}
\mu _{\mathrm{p}}=\frac{\left( \frac{\beta }{\theta _{\mathrm{E}}}\right)
^{2}+2}{2\left| \frac{\beta }{\theta _{\mathrm{E}}}\right| \sqrt{\left(
\frac{\beta }{\theta _{\mathrm{E}}}\right)^{2}+4}}+\frac{1}{2}\approx \frac{
\theta _{\mathrm{E}}}{2\left| \beta \right| },  \label{50a}
\end{equation}
\begin{equation}
\mu _{\mathrm{s}}=\frac{\left( \frac{\beta }{\theta _{\mathrm{E}}}\right)
^{2}+2}{2\left| \frac{\beta }{\theta _{\mathrm{E}}}\right| \sqrt{\left(
\frac{\beta }{\theta _{\mathrm{E}}}\right)^{2}+4}}-\frac{1}{2}\approx \frac{
\theta _{\mathrm{E}}}{2\left| \beta \right| }.  \label{50b}
\end{equation}
So, the total amplification $\mu _{\mathrm{wf}}$ of the weak field images is 
\begin{equation}
\mu _{\mathrm{wf}}=\mu _{\mathrm{p}}+\mu _{\mathrm{s}}\approx \frac{\theta _{
\mathrm{E}}}{\left| \beta \right| }=\frac{1}{\left| \beta \right| }\sqrt{
\frac{2d_{\mathrm{ls}}}{d_{\mathrm{os}}d_{\mathrm{ol}}}}.  \label{50c}
\end{equation}

Dividing Eq. (\ref{47b}) by (\ref{50c}) we have that
\begin{equation}
\frac{\mu _{\mathrm{R}}}{\mu _{\mathrm{wf}}}=\left( \frac{d_{\mathrm{os}}}{
d_{\mathrm{ls}}d_{\mathrm{ol}}}\right)^{\frac{3}{2}}f(q),  \label{51}
\end{equation}
where
\begin{equation}
f(q)=\frac{4D}{\sqrt{2}AB^{2}}\left( \frac{D}{B^{2}
} \frac{e^{-12\pi /A}}{1-e^{-8\pi /A}}+C\frac{e^{-6\pi /A}}{1-e^{-4\pi /A}}
\right).
\label{51b}
\end{equation}
f(q) takes values between $1\times 10^{-2}$ and $3\times 10^{-2}$,
for $|q|\in [0,0.5]$, so the relativistic images are much less
amplified than the weak field images. To give an example, the
prefactor multiplying $f(q)$ is $(\mathcal{D}/2M)^{-3/2}$, if the lens is
halfway between the observer and the source, $\mathcal{D} \approx D_{os}/4$
which, unless the black hole is very massive and or near the Earth
makes the amplification extremely small.

\section{An example: black hole in the galactic halo}

In this section we consider as a lens a black hole with mass
$M=7M_{\odot }$ , negligible angular momentum ($S\approx 0$) and
charge $Q$ placed at the galactic halo ($D_{\mathrm{ol}}=4
\mathrm{kpc}$) \cite{punsly}. The source is a star with radius
$R_{\mathrm{s}}=R_{\odot }$ located in the galactic
bulge ($D_{\mathrm{os}}=8 \mathrm{kpc}$). This provides a model for lensing 
by black hole candidates similar to the case discussed in Ref. 
\cite{Punsly:2000xb} and is  only intended to give some feeling of the 
numerical values implied in the strong field lensing of charged compact 
objects. \\

The positions and magnifications of the relativistic images for a
point source are given, respectively, by Eqs. (\ref{44a},
\ref{44b}) and (\ref{46}). The weak field images are formed near
the Einstein angular radius $\theta _{\mathrm{E}}\approx
\mathrm{2.669\,{mas}}$\textrm{\ (milliarc second)}. The
relativistic images have angular radius of the order of the
relativistic Einstein rings, the outer one is $\theta
_{1}^{\mathrm{E}}$ and the others approach to $\theta _{\infty
}^{\mathrm{E}}$. In Table 2 the angular positions of the horizon,
the photon sphere, the outer Einstein ring and the limiting value of
the Einstein rings are given for different values of $Q$.

\begin{table}[h]
\begin{center}
\caption{Angular positions of the horizon, the photon sphere,
the outer Einstein ring ($\theta _{1}^{E}$) and the limiting value of the Einstein rings ($\theta _{\infty }^{E}$) for different values of $Q$.}
\begin{tabular}{ccccc}\hline
$\left| Q\right| $ & $0$ & $0.1M$ & $0.5M$ & $M$ \\ \hline
$\theta _{H} \textrm{\ (mas)}$ & $3.455\times 10^{-8}$ & $3.447\times
10^{-8}$ & $3.224\times 10^{-8}$ & $1.727\times 10^{-8}$ \\
$\theta _{ps} \textrm{\ (mas)}$ & $5.181\times 10^{-8}$ &
$5.171\times 10^{-8}$ & $4.876\times 10^{-8}$ & $3.455\times 10^{-8}$ \\
$\theta _{1}^{E} \textrm{\ (mas)}$ & $8.987\times 10^{-8}$ & $8.973\times
10^{-8}$ & $8.595\times 10^{-8}$ & $6.957\times 10^{-8}$ \\
$\theta _{\infty }^{E}$ \textrm{\ (mas)} & $8.977\times 10^{-8}$ &
$8.960\times 10^{-8}$ & $8.581\times 10^{-8}$ & $6.910\times 10^{-8}$\\  \hline
\end{tabular}
\label{rnt2}
\end{center}
\end{table}

\begin{figure}[t!]
\vspace{-1.5cm}
\begin{center}
\includegraphics[width=7.1cm,height=10cm]{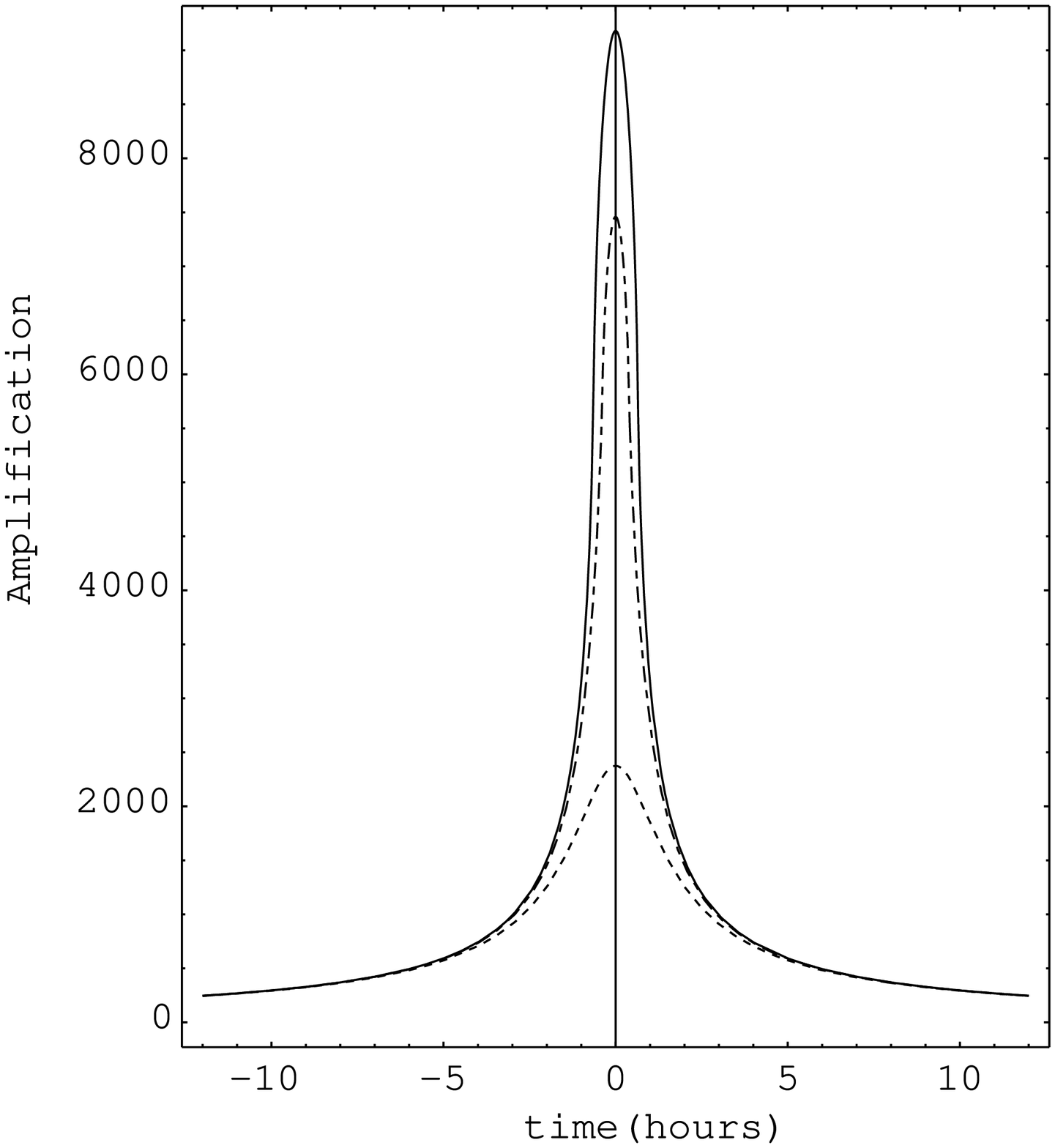} \hspace{0.cm}
\includegraphics[width=7.1cm,height=10cm]{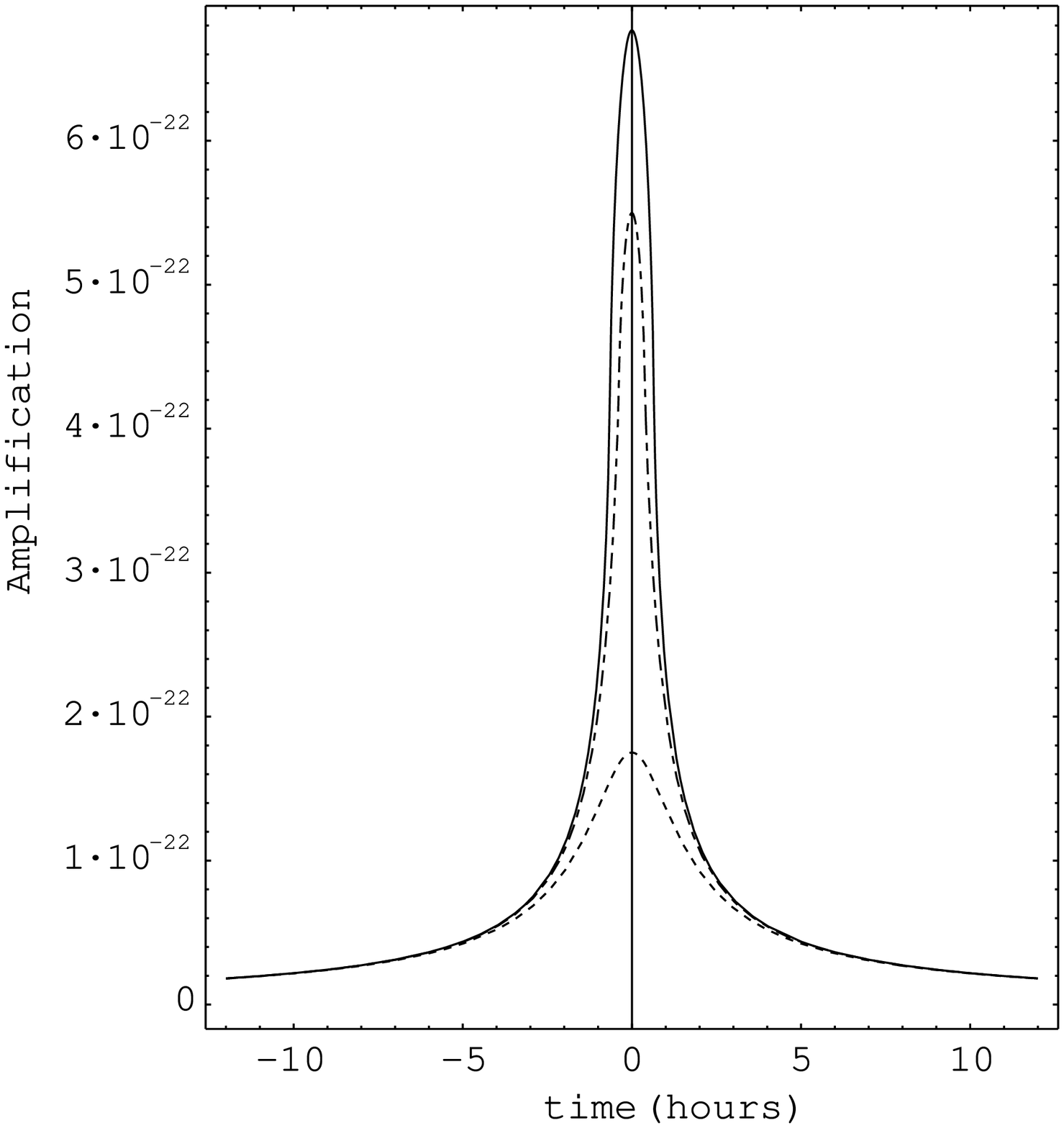} \\
\vspace{-2.5cm}
\includegraphics[width=7.1cm,height=10cm]{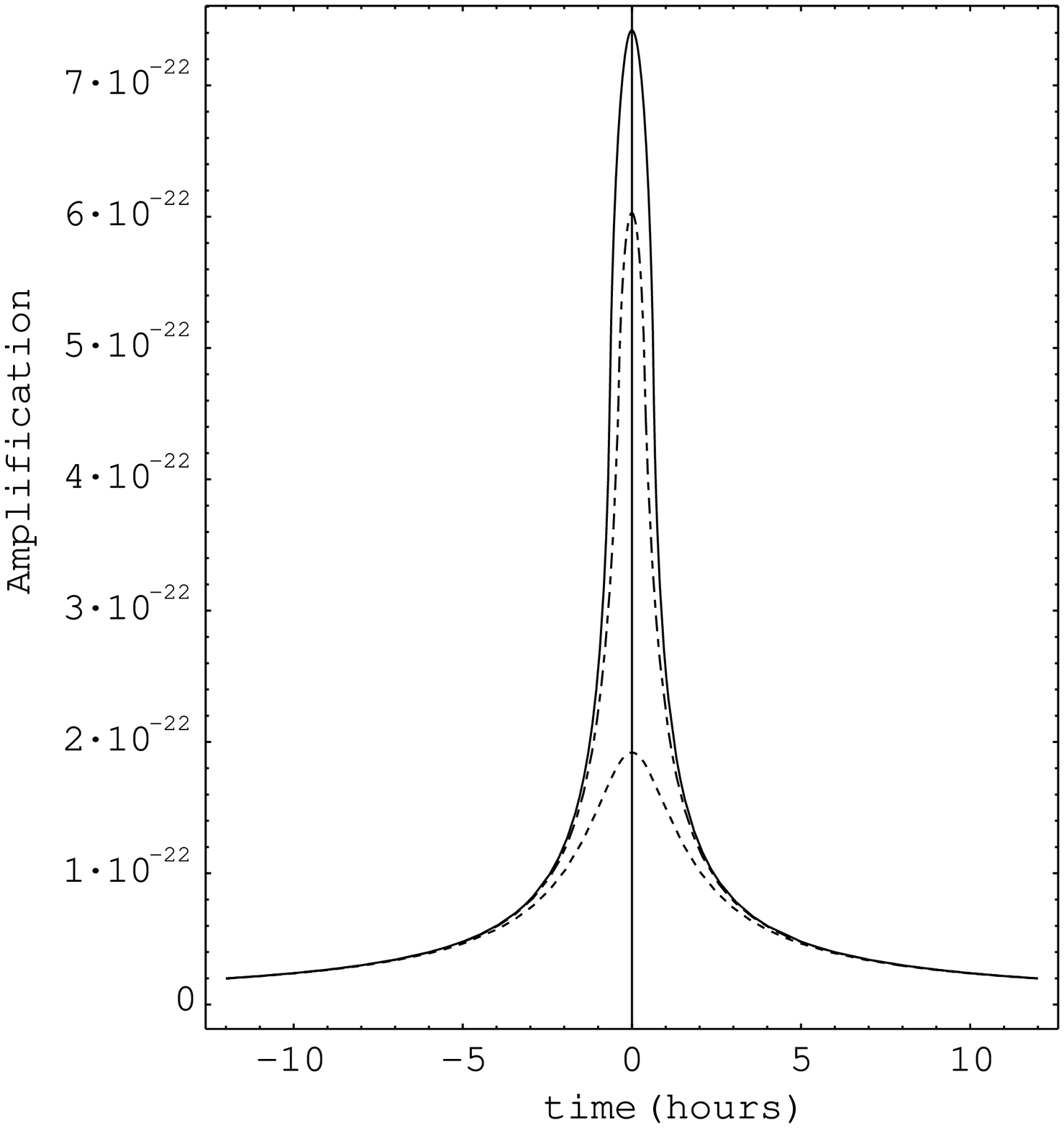} \hspace{0.cm}
\includegraphics[width=7.1cm,height=10cm]{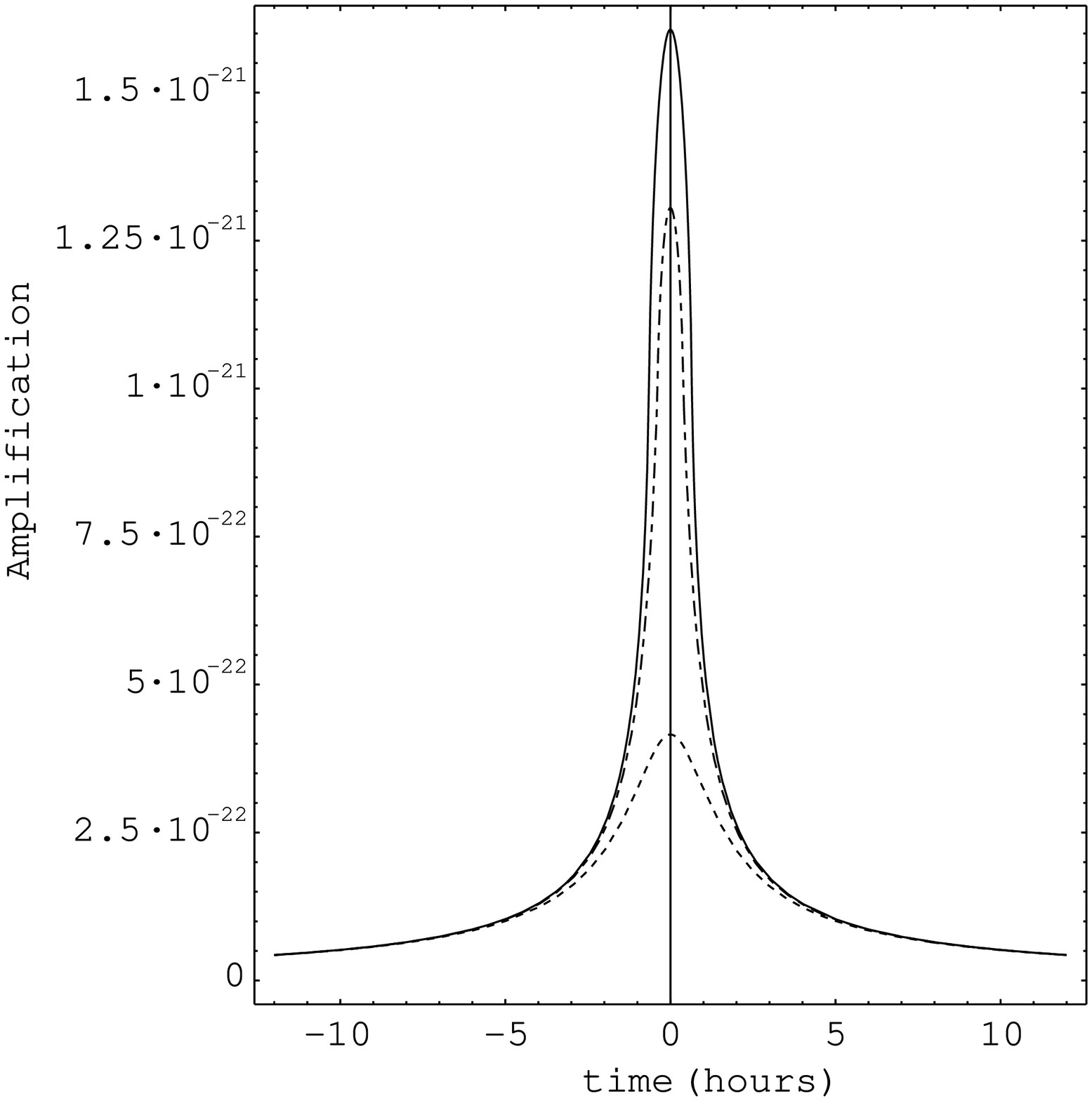}
\end{center}
\vspace{-1.5cm} \caption{Light curves for a lensing event by a
black hole with mass $ M=7M_{\odot }$ and charge $Q$ placed at the
galactic halo. The source is a star of radius
$R_{\mathrm{s}}=R_{\odot }$ placed at the galactic bulge. Upper
panel, left: total amplification of the primary and secondary
images. Upper panel, right: $Q=0$, total amplification of the
relativistic images. Lower panel, left: $|Q|=0.5\,M$, total
amplification of the relativistic images. Lower panel, right:
$|Q|=M$, total amplification of the relativistic images. In all
plots, the solid line corresponds to $\beta _{0}=0$, the dash-dot
line to $\beta _{0}=0.8\beta _{\mathrm{s}}$ and the dot line to
$\beta _{0}=2\beta _{\mathrm{s}}$.} \label{rnf5}
\end{figure}

The source is extended, so we have to integrate over its
luminosity profile to obtain the magnification:
\begin{equation}
\mu =\frac{\int\!\!\int_{S}\mathcal{I}\mu _{p}dS}{\int\!\!\int_{S}\mathcal{I}dS} \label{52a}
\end{equation}
where $\mathcal{I}$ is the surface intensity distribution of the source and
$\mu _{p}$ is the magnification corresponding to each point of the source.
We take the source as uniform, so
\begin{equation}
\mu =\frac{\int\!\!\int_{S}\mu _{p}dS}{\int\!\!\int_{S}dS}.
\label{52b}
\end{equation}
For both the relativistic and the weak field cases, the
amplifications for a point source are proportional to $1/\beta $:
\begin{equation}
\mu \propto I=\frac{\int\!\!\int_{S}\frac{1}{\beta }dS}{\int\!\!\int_{S}dS}. \label{52c}
\end{equation}
To obtain $I$ we use polar coordinates $(R,\varphi )$ in the source plane, with $R=0$ in the optical axis, and take the source as a disk $D(R_{\rm{C}},R_{\rm{S}})$ of radius $R_{\rm{S}}$ centered in $R_{\rm{C}}$. Then
\begin{equation}
I= \frac{\int\!\!\int_{D(R_{C},R_{S})}\frac{1}{\beta }RdRd\varphi }{
\pi R_{S}^{2}}, \label{52d}
\end{equation}
and, using that $\beta =R/D_{\rm{os}}$ is the angular position of each point of the source, we have
\begin{equation}
I=\frac{\int\!\!\int_{D(\beta _{C},\beta _{S})}\frac{1}{\beta }
\beta d\beta d\varphi }{\pi \beta _{S}^{2}}=\frac{\int\!\!\int_{D(\beta
_{C},\beta _{S})}d\beta d\varphi }{\pi \beta _{S}^{2}}. \label{52e}
\end{equation}
where $D(\beta _{\mathrm{c}},\beta _{\mathrm{s}})$ is the disk with angular
radius $\beta _{\mathrm{s}}=5.8\times 10^{-4}\mathrm{mas}$ centered in $\beta _{\mathrm{c}}$. The last integral can be calculated in terms of elliptic functions \cite{bozza}:
\begin{eqnarray}
I&=&\frac{2Sign[\beta _{\mathrm{s}
}-\beta _{\mathrm{c}}]}{\pi \beta _{\mathrm{s}}^{2}}\left[ (\beta _{\mathrm{s
}}-\beta _{\mathrm{c}})E\left( \frac{\pi }{2},-\frac{4\beta _{\mathrm{s}
}\beta _{\mathrm{c}}}{(\beta _{\mathrm{s}}-\beta _{\mathrm{c}})^{2}}\right)
+\right.  \nonumber \\
&&+\left. (\beta _{\mathrm{s}}+\beta _{\mathrm{c}})F\left( \frac{\pi }{2},-
\frac{4\beta _{\mathrm{s}}\beta _{\mathrm{c}}}{(\beta _{\mathrm{s}}-\beta _{
\mathrm{c}})^{2}}\right) \right] ,  \label{52}
\end{eqnarray}
where
\begin{equation}
F(\phi _{0},\lambda )=\int_{0}^{\phi _{0}}\left( 1-\lambda \sin ^{2}\phi
\right) ^{-\frac{1}{2}}d\phi ,  \label{53}
\end{equation}
\begin{equation}
E(\phi _{0},\lambda )=\int_{0}^{\phi _{0}}\left( 1-\lambda \sin ^{2}\phi
\right) ^{\frac{1}{2}}d\phi ,  \label{54}
\end{equation}
are elliptic integrals of the the first and second kind respectively, with
the arguments $\phi _{0}$ and $\lambda $.\\

In Fig. \ref{rnf5} the light curves for a transit event are shown.
The relative velocity between the lens and the source is $v=300\,
\mathrm{km/s}$. The angular position of the center of the source
is given by $\beta _{ \mathrm{c}}=\sqrt{\beta _{0}^{2}+\left(
\frac{vt}{D_{\mathrm{os}}}\right)^{2} }$, where $\beta _{0}$ is
the minimum angular position of the center of the source
corresponding to $t=0$. We see that the transit event takes only a
few hours, with great amplification for the weak field images and,
as expected, a negligible magnification for the relativistic
images. However, it is interesting to remark that with the
presence of charge the amplification of the relativistic images
increases respect to the Schwarzschild case. For a charge $|Q|=0.5
M$ we have about $10$ \% more amplification than in a similar
black hole with no charge. In the case $ |Q|=1 M$ the total
amplification of the relativistic images is about $140$ \%
 stronger than the corresponding value for a Schwarzschild black
hole of the same mass.

\section{Conclusions}

In this paper we developed the formalism for charged black hole
lensing. We find the positions and magnifications of the
relativistic images within the strong field limit. The
relativistic images are more prominent when there is a close
alignment between the observer, the lens, and the source. They
are, though, very faint, even with complete alignment, but much
stronger than in the case of Schwarzschild black holes for large
charges. The angular separation of the relativistic images is
beyond the angular resolution of current optical technologies. For
local black holes (e.g. in the Gould Belt, a star forming region
at 600 pc), however, if radio or X-ray background sources can be
used, the angular separations of $\mu as$ could be resolved with
space techniques.

The NASA Constellation-X \cite{consx} mission, to be launched in
2008, is optimized to study the iron K$\alpha$ line features and
will determine the black hole mass and spin for a large number of
systems. Still, Constellation-X will provide an indirect measure
of the properties of the region within a few event horizon radii.
NASA-planned MAXIM mission \cite{maxim}, a $\mu$-arcsec X-ray
imaging mission, will be able to take direct X-ray pictures of
regions of the size of a black hole event horizon. Both of these
space mission will have the ability to give us proofs of black
hole existence, and possibly, to distinguish even among different
black hole solutions. The project ARISE (Advanced Radio
Interferometry between Space and Earth) will use the technique of
Space VLBI. The mission, to be launched in 2008, will be based on
a 25-meter inflatable space radio telescope working between 8 and
86 GHz \cite{ulv99}. ARISE will provide resolution of 15
$\mu$arc-seconds or better, 5-10 times better than that achievable
on the ground. At frequencies of 43 and 86 GHz, this would imply
resolution of light weeks to light months in distant quasars and
will complement the gamma-ray and X-ray observations. ARISE, then,
could also study gravitational lenses at resolutions of tens of
$\mu$arc-seconds, and as such, could prove important in the
detection of charged black holes. However, it is not only
resolution what is needed in order to detect the relativistic
images. The main problem will be that they are very much
demagnified. They would certainly pose a challenge to
observations, if ever possible, enhanced by the fact that the
relativistic images are closed to the non-relativistic ones, which
are much more intense.

\section*{Acknowledgements}

We gratefully acknowledge discussions with Salvatore Capozziello
and Valerio Bozza (U. of Salerno, Italy). This work has been
partially supported by UBA (UBACYT X-143, EFE), CONICET (DFT, and
PIP 0430/98, GER), ANPCT (PICT 98 No. 03-04881, GER), and
Fundaci\'{o}n Antorchas (separate grants to GER and DFT).

\section*{Appendix: Deflection angle in terms of elliptic integrals}

It is convenient for the calculations to express the integrals of Eqs. (\ref
{10}) and (\ref{24}) in terms of standard elliptic integrals.\\
Eq. (\ref{10}) can be put in the form
\begin{equation}
\alpha (x_{0})=\frac{2x_{0}}{\sqrt{1-\frac{1}{x_{0}}+\frac{q^{2}}{x_{0}^{2}}}
}\int_{x_{0}}^{\infty }\frac{dx}{\sqrt{P(x)}}-\pi ,  \label{a1}
\end{equation}
where
\begin{equation}
P(x)=x^{4}+x_{0}^{2}\left( 1-\frac{1}{x_{0}}+\frac{q^{2}}{x_{0}^{2}}\right)
^{-1}(-x^{2}+x-q^{2}),  \label{a2}
\end{equation}
is a fourth degree polynomial. The four roots of $P(x)$ depend only on $
x_{0} $ and $q$ and they are real for $x_{0}>x_{\mathrm{ps}}$ and $\left|
Q\right| \leq M$. Calling these roots $r_{1}>r_{2}>r_{3}>r_{4}$ , the first
one is $r_{1}=x_{0} $ and the others have long expressions (not given here
for lack of space) that can be obtained with standard software. Then we can
write the exact integral of Eq. (\ref{10}) as
\begin{equation}
\alpha (x_{0})=G(x_{0})F(\phi _{0},\lambda )-\pi ,  \label{a3}
\end{equation}
with
\begin{equation}
G(x_{0})=\frac{4x_{0}}{\sqrt{1-\frac{1}{x_{0}}+\frac{q^{2}}{x_{0}^{2}}}}
\frac{1}{\sqrt{(r_{1}-r_{3})(r_{2}-r_{4})}},  \label{a4}
\end{equation}
and
\begin{equation}
F(\phi _{0},\lambda )=\int_{0}^{\phi _{0}}\left( 1-\lambda \sin ^{2}\phi
\right) ^{-\frac{1}{2}}d\phi ,  \label{a5}
\end{equation}
an elliptic integral of the first kind with arguments
\begin{equation}
\phi _{0}=\arcsin \sqrt{\frac{(r_{2}-r_{4})}{(r_{1}-r_{4})}},  \label{a6}
\end{equation}
\begin{equation}
\lambda =\frac{(r_{1}-r_{4})(r_{2}-r_{3})}{(r_{1}-r_{3})(r_{2}-r_{4})}.
\label{a7}
\end{equation}
The derivative of the deflection angle can be expressed:
\begin{equation}
\frac{d\alpha }{dx_{0}}=F(\phi _{0},\lambda )\frac{dG}{dx_{0}}+G(x_{0})\frac{
\partial F}{\partial \lambda }\frac{d\lambda }{dx_{0}}+G(x_{0})\frac{
\partial F}{\partial \phi _{0}}\frac{d\phi _{0}}{dx_{0}},  \label{a10}
\end{equation}
and can be written as
\begin{eqnarray}
\frac{d\alpha }{dx_{0}} & = & F(\phi _{0},\lambda )\frac{dG}{dx_{0}}
+G(x_{0})\left( -\frac{E(\phi _{0},\lambda )}{2(\lambda -1)\lambda }-\frac{
F(\phi _{0},\lambda )}{2\lambda }+\frac{\sin (2\phi _{0})}{4(\lambda -1)
\sqrt{1-\lambda \sin ^{2}\phi _{0}}}\right) \frac{d\lambda }{dx_{0}}+
\nonumber \\
& & +\frac{G(x_{0})}{\sqrt{1-\lambda \sin ^{2}\phi _{0}}}\frac{d\phi _{0}}{
dx_{0}},  \label{a11}
\end{eqnarray}
where
\begin{equation}
E(\phi _{0},\lambda )=\int_{0}^{\phi _{0}}\left( 1-\lambda \sin ^{2}\phi
\right) ^{\frac{1}{2}}d\phi ,  \label{a12}
\end{equation}
is an elliptic integral of the second kind with the arguments $\phi _{0}$
and $\lambda $ given above.

\end{document}